# Roughening dynamics of spontaneous radial imbibition

Yong-Jun Chen[1, 2,*], Kenichi Yoshikawa[2,*]

[1]Department of Physics, Shaoxing University, Shaoxing, Zhejiang Province 312000, China

[2]Faculty of Life and Medical Science, Doshisha University, Kyotanabe, Kyoto 610-394, Japan

∗Email address: chenyongjun@usx.edu.cn and keyoshik@mail.doshisha.ac.jp

**Abstract**

We performed an experimental observation on the spontaneous imbibition of water in a porous media in a radial Hele-Shaw cell and confirmed Washburn's law, $r \sim t^{1/2}$, where r is distance and t is time. Spontaneous imbibition with a radial interface window followed scaling dynamics when the front invaded into the porous media. We found a growth exponent (β=0.6) that was independent of the pressure applied at the liquid inlet. The roughness exponent decreased with an increase in pressure. The roughening dynamics of two dimensional spontaneous radial imbibition obey Family-Vicsek scaling, which is different from that with a one-dimensional planar interface window.

# 1. Introduction

Spontaneous imbibition is the displacement of a non-wetting fluid by a wetting fluid in a porous medium due to capillary force. It can be found not only in many everyday processes, for example, dipping a biscuit into water, cleaning the floor with a cloth, and the germination of a dry seed, but also in industrial processes, ranging from oil recovery to food processing [1-3]. The imbibition process is dominated by capillary pressure and viscous drag and is affected by quenched noise in the field of the porous medium. The roughening dynamics of the propagating invasion front are of fundamental interest [1, 2] and have been studied extensively, both theoretically and experimentally [1-5]. The statistical description of the roughening front has been linked to that of fire fronts, cracks and rupture lines, and so on, the roughening dynamics of which can be described using the Kardar-Parisi-Zhang (KPZ) equation [2]. The roughening dynamics of an imbibition front demonstrates scaling behavior [1, 2]. Various scaling behaviors (indicated by the growth exponent $\beta$ and the roughness exponent $\alpha$) have been reported experimentally [1]. The scaling behavior of an imbibition front is related to many factors, such as porosity, prewetting, gravity, disorder, and so on [1, 2]. Anomalous scaling dynamics of spontaneous imbibition have been reported in a Hele-Shaw cell [4]. However, there is insufficient agreement between theory and experimental results and our understanding of the fundamental mechanism of the roughening dynamics of spontaneous imbibition is still unclear [1]. Recently, the geometric effect of a propagating interface on roughening dynamics has attracted attention [6-10]. Non-Euclidian Geometry has been shown to affect the scaling dynamics of an imbibition front [6]. Many systems, such as solidification in crystal nucleation and bacterial cloning, exhibit radial propagation of the front interface. There is some controversy regarding whether the established scaling analysis is valid in systems with a different propagation window (for example, radial imbibition) [6-10]. One exceptional example is tumor growth. The traditional scaling method with a planar window has been adopted to determine the universality class of tumor evolution and is still a matter of debate [6-9]. The geometric role of the growth window is still not clear. Here, we observed the spontaneous radial imbibition in a porous medium within a Hele-Shaw cell. We found that spontaneous radial imbibition showed scaling dynamics that differed from those of one-dimensional planar imbibition in a Hele-Shaw cell.

## 2. Experiment

The experiment was performed using a Hele-Shaw cell constructed from two glass plates (30cm×30cm) that were aligned parallel to each other with a gap size of 0.5mm. The Hele-Shaw cell was filled with approximately two layers of glass beads with a diameter of 0.20±0.05mm. The glass beads were packed tightly by manual vibration during filling. Three of the four boundaries of the Hele-Shaw cell were sealed while one was left open so that air could come out. The inlet hole for fluid at the center of Hele-Shaw cell was connected to a reservoir of water by a tube. The water penetrated into the porous medium from the center of the Hele-Shaw cell. To change the inlet pressure the water, the water column was fixed at different heights from 0 to 12cm, as shown in Fig. 1. The invasion of the imbibition front was monitored using a digital CCD camera. Lateral optical illumination yielded a sharp contrast at the invading front. The experimental data were analyzed using image-analysis software. All experiments were performed under ambient conditions (24℃).

## 3. Results

We studies spontaneous imbibition with an increase in the water-column height (H), corresponding to an increase in the applied pressure difference at the fluid inlet. Figure 2 shows the morphology of the invasion front at various pressure differences (represented by the height of water column, H). The interface undergoes a roughening transition during imbibition into the porous medium of glass beads. The roughening process depends on the pressure difference at the inlet of fluid. Figure 3 shows Washburn's law for the average position $\bar{r}$ of the invasion front, $\bar{r} = At^{1/2}$, where $t$ is time and $A$ is a constant. According to Darcy's law, the value of $A^2$ should depend linearly on the pressure difference [4]. However, as shown in the inset of Fig. 3, the value of $A^2$ does not increase linearly as the height of the water column increases.

We digitized the invasion front and analyzed it numerically. The roughness of the front is the root-mean-square fluctuation

$$W = \langle r^2 - \bar{r}^2 \rangle^{1/2},$$

where $r$ and $\bar{r}$ are the radius to the center of mass (CM) and its mean value, and $\langle \rangle$ represents the average over the space. Figure 4 shows the log-log plot of roughness against time for different heights of the water column. The evolution of roughness exhibits scaling dynamics for interface propagation, $W \sim t^\beta$. The fluctuation in the log-log plot of the scaling law is attributable to the overhang and coalescence of the front. By fitting the data linearly, we obtain the growth exponent $\beta = 0.60 \pm 0.05$. The growth exponent does not depend on the pressure difference, as shown in Fig. 4. When the pressure difference is high (H=12.0cm), the data become somewhat scattered (Fig. 4).

To obtain the roughness exponent, we calculated the structure function of the invading front, $S(k,t) = |r(k,t)|^2$, where $r(k,t)$ is a Fourier transformation of the radius $r$ and $k$ is the wave number. For a self-affine interface, the structure function has the scaling form [11]

$$S(k,t) = k^{-(2\alpha+1)} f_s(kt^{1/z}) \text{ with}$$

$$f_s(u) \sim \begin{cases} u^{2(\alpha-\alpha_s)} & \text{if } u \gg 1 \\ u^{2\alpha+1} & \text{if } u \ll 1 \end{cases}$$

where $f_s(u)$ is a spectral scaling function, $\alpha$ is a roughness exponent, and $\alpha_s$ is a spectral roughness exponent. Figure 5 shows a plot of structure functions. From Fig. 5(a), we can obtain the roughness exponent $\alpha = 1.25$ when the pressure difference H=0, indicating that the invasion front is super-rough [11]. The collapse of the log-log plot of $S(k,t)k^{(2\alpha+1)}$ versus $kt^{1/z}$ with roughness exponent $\alpha = 1.25$ when H=0 is shown in the inset of Fig. 5(a). From this collapse, we find that $\alpha = \alpha_s$. Figure 5(b) shows the dependence of the roughness exponent on the pressure difference. A greater height H leads to a smoother front with a smaller value of the roughness exponent. We obtained roughness exponents of 1.10, 1.0, 0.93, and 0.75 for the height H values of 1.0cm, 4.0cm, 6.0cm, and 12.0cm, respectively. Thus, there is a clear trend in the roughening process of the front, as shown in Fig. 2.

## 4. Discussion

In one dimension with a planar interface window, anomalous roughening dynamics have been reported for spontaneous imbibition in a porous medium in a Hele-Shaw cell [4], and the roughening behavior depends on the pressure difference with a crossover from negative pressure to positive pressure [12]. We examined the evolution of local roughness and local structure factor in the present study on spontaneous radial imbibition. There was no apparent anomalous scaling. The roughness process followed Family-Vicsek (FV) scaling, as shown in the collapse of data in the inset of Fig. 5(a) using a dynamic exponent $z = \alpha/\beta = 2.08$ for H=0. This result is surprising. The invasion front should experience a similar roughening process due to the quenched disorder of the porous medium at a local point in both one-dimensional imbibition and two-dimensional radial imbibition. In addition, spontaneous radial imbibition has a growth exponent $\beta = 0.60$, independent of the pressure difference, while in one-dimensional imbibition, the value of growth exponent decreases when the pressure difference increases from negative to positive, as reported by Planet *et al.* [12]. At an intermediate pressure difference between a negative and positive pressure difference, Planet *et al.* [12] suggested that there exists a crossover regime which lacks scaling behavior. It has been reported that the roughness exponent 0.81 for small length-scales crosses over to quenched KPZ (QKPZ) roughness exponent 0.6 for spontaneous one-dimensional imbibition [13]. We can see that the roughening dynamics of spontaneous radial imbibition are markedly different from those of spontaneous one-dimensional imbibition with a planar interface window. We should note that the pressure difference in our experiment is different from that used by Planet *et al.*[12]. In our experiment, the pressure difference at the inlet of the fluid ranged from zero to over 10 centimeters, while in Planet's experiment the largest pressure difference was 1.5cm and, in fact, a negative pressure difference was possible. When H=1.5cm, the interface becomes smooth in one-dimensional imbibition. Based on the roughness exponents in our experiment, the interface is still rough even when the pressure difference H is over 6.0cm. These differences suggest that the mechanism of interface roughening should be different in the two cases.

Flow in a porous medium can be described using Darcy's Law, $\vec{\upsilon} = -\kappa \nabla P$, where $\vec{\upsilon}$ is velocity, P is the pressure and $\kappa$ is a constant. It has been shown that the interface roughening of spontaneous one-dimensional imbibition in a disordered medium can be characterized using the QKPZ equation [13, 14]. For spontaneous two-dimensional radial imbibition, the growth of the interface is generally described using a differential equation that is invariant under reparameterization [15]

$$\frac{\partial \vec{r}}{\partial t} = \nu \frac{1}{\sqrt{g}} \frac{\partial}{\partial s} \frac{1}{\sqrt{g}} \frac{\partial}{\partial s} \vec{r} + \bar{\upsilon}\hat{n} + \eta \qquad (1)$$

where $\vec{r}$ is a generalized coordinate on the front, t is time, s is the arc length on the moving front, $\bar{\upsilon}$ is the average velocity, $\hat{n}$ is normal to the moving front, $\nu$ is a constant, $\eta$ is the quenched noise and $g = |d\vec{r}/ds|^2$. Equation (1) can be reduced to the KPZ equation. If we set $g = 1.0$, we obtain

$$\frac{\partial \vec{r}}{\partial t} = \nu \frac{\partial^2}{\partial s^2} \vec{r} + \upsilon\hat{n} + \eta \qquad (2)$$

Such growth should be different from that with a KPZ-like equation using the expression $r(\theta, t)$ without overhang [16]. As demonstrated by Singha [16], the geometry of the interface affects the auto-correlation exponent and thus the growth exponent. We implement the Eq. 2 numerically. Figure 6 shows the numerical results. The growth exponent is 0.58, 0.60, 0.56, 0.57 and 0.58 for averaged velocities of 0.1, 0.4, 0.8, 1.2 and 1.6, respectively. The value of the exponent approximately agrees with the experimental measurement ($\beta = 0.60$). A plot of the structure function S(k,t) against wavenumber shows the dependence of the roughness exponent on the average velocity (Fig. 6(b)). The correspondence of the average velocity in Eq. 2 to the actual velocity of the imbibition front in the experiment is not clear. The roughness exponents for a relatively large wavenumber in Fig. 6(b) are close to 1.25, which is the roughness exponent for H=0 in the experiment. However, Eq. 2 still has a drawback; it does not indicate the anisotropy of the pinning effect on the evolution of the front. As shown in Fig. 2, anisotropy of the pinning from the quenched disorder of the porous medium is obvious and dominates the morphology of the invasion front. The anisotropic pinning affects the invasion not only locally but also globally. Thus, the pinning-depinning process in imbibition dominated the roughening process of the front. An overhang is ubiquitous in imbibition with a low pressure difference. The viscosity of the liquid also has an effect on the morphology of the front [17]. For imbibition with a higher pressure difference, Eq. 2 is not valid. In this case, the imbibition of water in the porous medium is dominated by both capillary force and the external pressure difference, and is similar to forced fluid imbibition [18]. Both effects should be considered.

During imbibition, the porous medium is not saturated with water and waterless voids can be found behind the invasion front. The pressure of the water will affect the degree of water saturation in the porous medium. Thus, an increase in the pressure difference (height of the water column) increases not only the velocity of front propagation, but also the degree of saturation of water in the porous medium. This effect can be expected to influence on the roughening behavior of the front. In particular, a second imbibition has been found when the pressure difference is relatively large. Figure 7(a), (b) shows the second imbibition after the first imbibition discussed above. During the first imbibition, the porous medium of glass beads is not saturated with water. In the first imbibition, water only sweeps through the porous medium but does not fill all of the space between the glass beads. The air-water interface in the imbibed area is pinned and forms voids. However, when the pressure difference is rather high, the pinned air-water interface is depinned. The porous medium becomes saturated with water and the second imbibition appears. The second imbibition strongly influences on the propagation of the first imbibition. Under certain conditions, the first imbibition almost stops as observed with a pressure difference of 6cm (Fig.7), and thus does not obey Washburn's law. While Washburn's law was also seen with a water column of H=12cm, the scaling plot is scattered, as shown in Fig. 4. The effect of the second imbibition on the first imbibition is still not clear. The second imbibition does not follow the Washburn's law, as shown by Fig. 7(c). The average position of the imbibition front shows scaling behavior against time, but the value of exponent deviates from 0.5. The roughening dynamics follow scaling behavior as a whole, but the data are scattered.

**5. Conclusion**

　　We have demonstrated the spontaneous imbibition of water in a porous medium confined in a Hele-Shaw cell. Washburn's law was verified, which indicates that Washburn's law is not affected by the geometric shape of the imbibition front. A common growth exponent was observed for various pressure differences. As the pressure difference at the inlet of the fluid is increased, the value of roughness exponent decreases. The roughening dynamics of spontaneous radial imbibition that obeys FV scaling are not identical to the anomalous scaling dynamics of one-dimensional imbibition with a planar window. Thus, it has been made clear that geometric shape of the interface will affect the exponent of roughening dynamics. We cannot take for granted that traditional scaling analysis of roughening dynamics with one-dimensional planar window is valid for circular interface growing. And it is misleading to assume that one-dimensional growth and radial growth of the same interface will belong to the same universality class. We hope that our findings will provide new insight for studies on the roughening dynamics of an interface with a radial window.

**Acknowledgement**

This work was supported by a grant from the National Natural Science Foundation of China (No. 11204181) and by SRF for ROCS, SEM. We also wish to acknowledge MEXT KAKENHI Grants-in-Aid (Nos. 22654046 and 25103012).

Figure captions

Fig. 1 Experimental setup and a typical propagating front in spontaneous imbibition. (a) Experimental setup. (b) A typical invasion front. Scale bar: 20mm.

Fig. 2 Spatial-temporal propagation of the imbibition front at different pressure differences (color online). (a) H=0cm, (b) H=1.0cm, (c) H=4.0cm, (d) H=6.0cm, (e) H=12.0cm.

Fig. 3 Plot of the squared mean radius against time. The linear fit shows Washburn's law for the imbibition of water in the porous medium. The inset is a plot of the squared prefactor $A^2$ in the Washburn equation against the height of the liquid reservoir.

Fig. 4 Log-log plot of roughness against time for various pressure differences. The slopes of the dashed fitting lines are 0.60 for H=0cm, 1.0cm, 4.0cm, and 6.0cm. In contrast, the slope of the fitting linefor H=12.0cm is 0.50.

Fig. 5 Structure function. (a) Structure function of fronts at different time when H=0cm. The inset is the collapse of the log-log plot of $S(k,t)k^{(2\alpha+1)}$ versus $kt^{1/z}$ ($z=\alpha/\beta=2.08$). (b) Structure function for invasion fronts at various pressure differences.

Fig. 6 Numerical result. (a) Log-log plot of width W against time. (b) Log-log plot of structure function $S(k,t)$ against wavenumber $k$. The slopes of the dashed lines are indicated in the figure. The parameters used for the simulation were $v=-4.0$ and time step $\Delta t=0.0002$. The value of $\upsilon$ is shown.

Fig. 7 Propagation of the second imbibition front (color online). (a) A typical second imbibition front. Scale bar is 50mm . (b) Spatial-temporal evolution of the second imbibition front. (c) Log-log plot of the roughness of the interface and radius against time. The pressure difference is H=6cm.

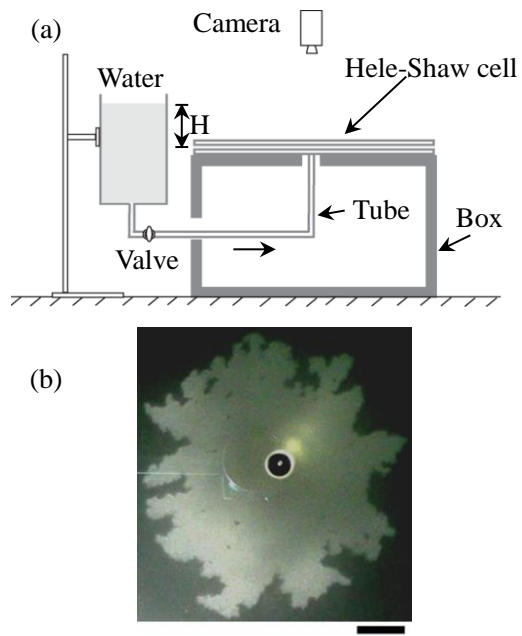

Figure 1 Yong-Jun Chen, Kenichi Yoshikawa

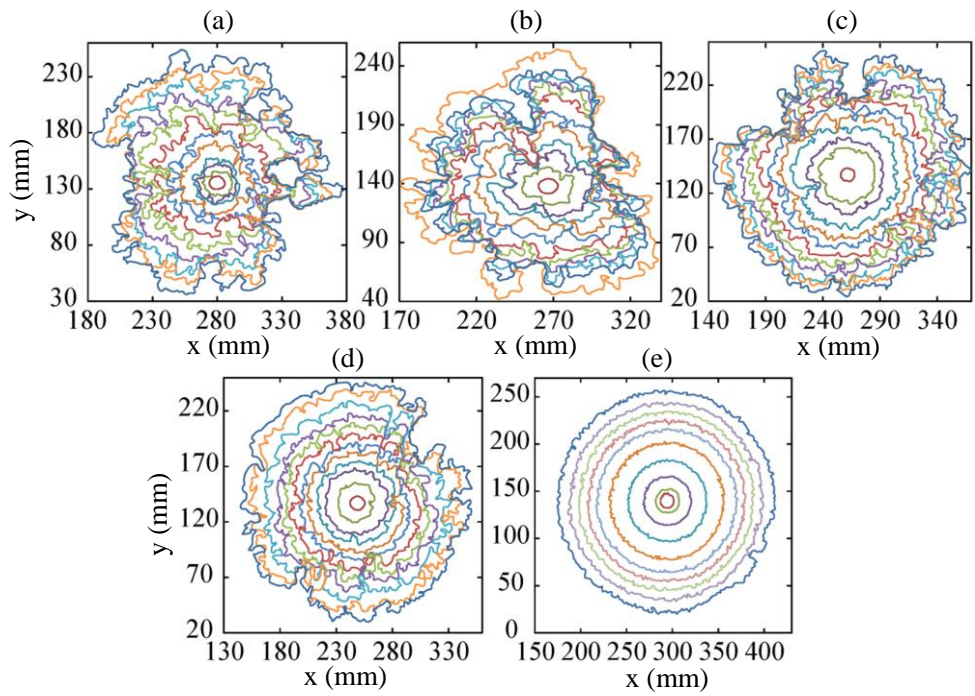

Figure 2 Yong-Jun Chen, Kenichi Yoshikawa

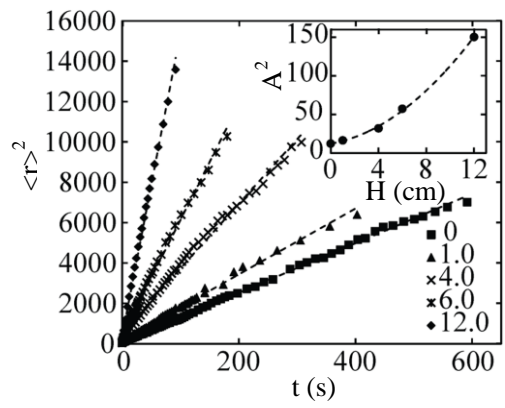

Figure 3 Yong-Jun Chen, Kenichi Yoshikawa

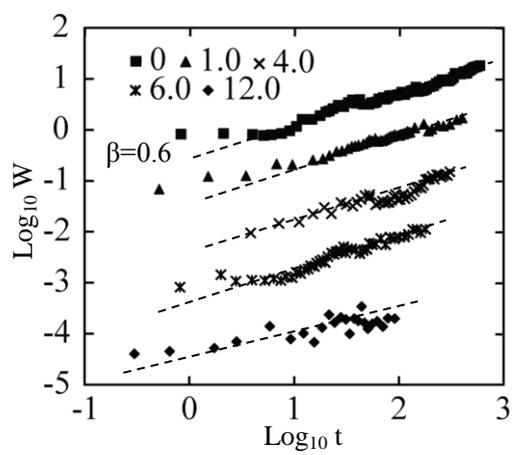

Figure 4 Yong-Jun Chen, Kenichi Yoshikawa

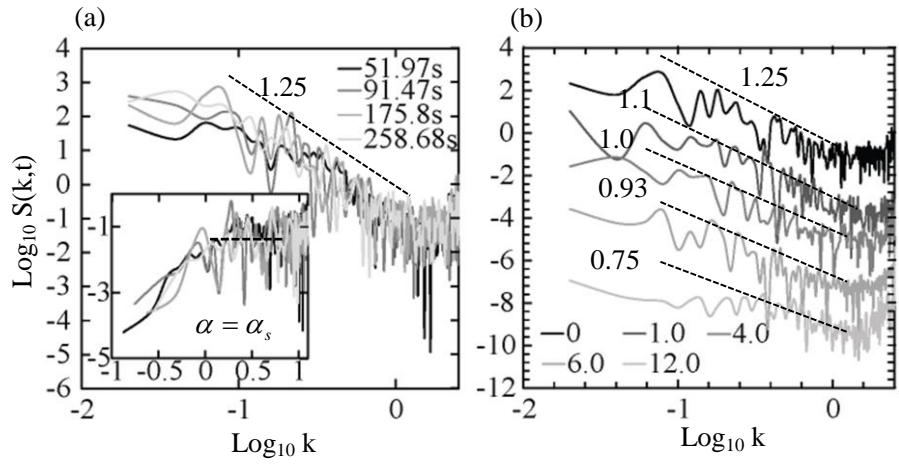

Figure 5 Yong-Jun Chen, Kenichi Yoshikawa

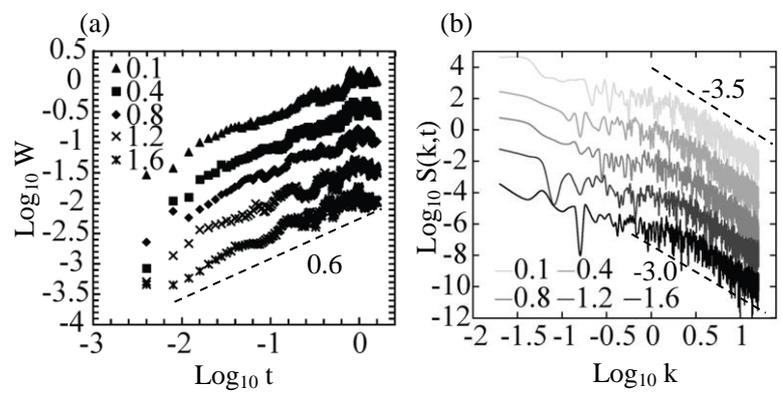

Figure 6 Yong-Jun Chen, Kenichi Yoshikawa

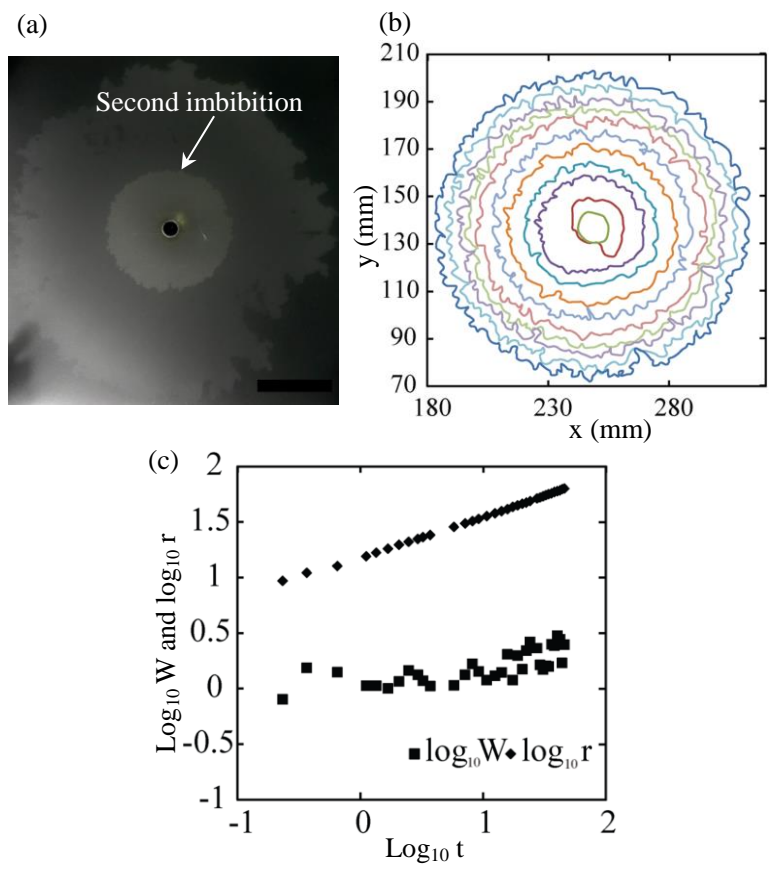

Figure 7 Yong-Jun Chen, Kenichi Yoshikawa